\documentclass[prl,showpacs,floatfix,twocolumn]{revtex4}
\usepackage{amsfonts}
\usepackage{stmaryrd}
\usepackage{bbm}
\usepackage{mathrsfs}
\usepackage{tipa}
\usepackage{amssymb}
\usepackage{txfonts}
\usepackage{graphicx}
\usepackage{dcolumn}
\usepackage{epstopdf}
\usepackage{array}

\newcommand{\PreserveBackslash}[1]{\let\temp=\\#1\let\\=\temp}
\newcolumntype{C}[1]{>{\PreserveBackslash\centering}p{#1}}
\newcolumntype{R}[1]{>{\PreserveBackslash\raggedleft}p{#1}}
\newcolumntype{L}[1]{>{\PreserveBackslash\raggedright}p{#1}}

\begin{document}

\newcommand*{\cm}{cm$^{-1}$\,}


\title{Revealing multiple density wave orders in non-superconducting titanium oxypnictide Na$_2$Ti$_2$As$_2$O}

\author{Y. Huang}
\author{H. P. Wang}
\author{R. Y. Chen}
\author{X. Zhang}
\author{P. Zheng}
\author{Y. G. Shi}
\author{N. L. Wang}

\affiliation{Beijing National Laboratory for Condensed Matter
Physics, Institute of Physics, Chinese Academy of Sciences,
Beijing 100190, China}
%


\begin{abstract}

We report an optical spectroscopy study on the single crystal of Na$_2$Ti$_2$As$_2$O, a sister compound of superconductor BaTi$_2$Sb$_2$O. The study reveals unexpectedly two density wave phase transitions. The first transition at 320 K results in the formation of a large energy gap and removes most part of the Fermi surfaces. But the compound remains metallic with residual itinerant carriers. Below 42 K, another density wave phase transition with smaller energy gap scale occurs and drives the compound into semiconducting ground state. These experiments thus enable us to shed light on the complex electronic structure in the titanium oxypnictides.

\end{abstract}

\pacs{71.45.Lr, 75.30.Fv, 74.25.Gz, 74.70.-b}

\maketitle

The low temperature broken-symmetry states in low-dimension materials, such as superconductivity, charge/spin density wave (CDW/SDW), are among the most fascinating collective phenomena in solids and the interplay between them has been a subject of considerable interest in condensed matter physics \cite{densitywave,Wilson1,Wilson2,Mongeau,Becker,Degiorgi}. Most density wave (DW) instabilities (either CDW or SDW) are driven by the nesting topology of Fermi surfaces (FSs), that is, the matching of sections of FS to others by a wave vector 2\textbf{k}$_F$, where the electronic susceptibility has a divergence. Brought about by electron-phonon or electron-electron interactions, a single particle energy gap opens in the nested regions of FSs, resulting in the lowering of the electronic energy of the system. In one-dimensional (1D) system the DW phase transition usually causes an semiconducting ground state due to the opening of a full energy gap arising from the perfect nesting of Fermi surfaces. However, for two-dimensional (2D) or three-dimensional (3D) materials, the CDW or SDW ground states mostly remain metallic due to the formation of partial energy gap induced by the imperfect nesting of FSs. To date, there seems no reported example of a truly semiconducting ground state caused by the DW phase transition in a 2D or 3D system.

Recently, a new superconducting system Ba$_{1-x}$Na$_x$Ti$_2$Sb$_2$O (T$_c$$\sim$2-5 K) has attracted much attention \cite{Yajima,Doan,Zhai}. The system belongs to a two-dimensional (2D) titanium oxypnictide family, consisting of
alternate stacking of conducting octahedral layers Ti$_2$Pn$_2$O (Pn=As, Sb) and other insulating layers (e.g. Na$_2$,
Ba, (SrF)$_2$, (SmO)$_2$) \cite{Adam,Axtell,Ozawa1,Ozawa2,Ozawa3,Ozawa4,Liu1,Wang,Liu2,Zhai}. The undoped compounds in this family commonly exhibit phase transitions below certain temperatures (e.g. 114 K for Na$_2$Ti$_2$Sb$_2$O \cite{Ozawa1}, 320 K for Na$_2$Ti$_2$As$_2$O \cite{Axtell,Ozawa1,Ozawa2,Ozawa3,Liu1}, 45 K for BaTi$_2$Sb$_2$O \cite{Yajima}, 200 K for BaTi$_2$As$_2$O \cite{Wang}), as characterized by the sharp jumps in resistivity and drops in magnetic susceptibility. First principle band structure calculations indicate that the phase transitions are driven by the DW instabilities arising from the nested electron and hole FSs \cite{Pickett,Biani,Singh,Yan,Subedi}. As superconductivity emerges only in compound with low phase transition temperature and T$_c$ is further enhanced when the phase transition temperature was suppressed by doping, the family offers a new playground to study the interplay between superconductivity and DW instabilities. Understanding the electronic properties in the undoped compounds is an essential step towards understanding the superconductivity in this family. Among the different members in the family, Na$_2$Ti$_2$As$_2$O appears to be particularly interesting. The first principle calculations indicate that the compound has an SDW ground state with a blocked checkerboard antiferromagnetic (AFM) order, and the ordered state is not metallic but semiconducting with an energy gap of 0.15 eV \cite{Yan}. Up to the present, there have been little experimental studies on the material. It is highly desirable to probe its electronic properties and, in particular, to examine whether such an semiconducting ground state is really realized in Na$_2$Ti$_2$As$_2$O system.

In this work we present the optical spectroscopic study on Na$_2$Ti$_2$As$_2$O single crystal. Unexpectedly, the study reveals two DW phase transitions for the compound. The first one at high temperature ($\sim$320 K) results in the formation of a large DW energy gap with $2\Delta$=2600 \cm and removes most parts of the FSs. But the compound remains a metallic nature after this phase transition. Below 42 K, another DW gap forms and removes the rest of the FSs totally. Such two steps phase transitions, which eventually lead to a semiconducting ground state, were not anticipated from the first principle calculation. Na$_2$Ti$_2$As$_2$O manifests much more complex than one would expect. It is likely to be the first proven example of a quasi-2D material undergoing metal-insulator (M-I) transition driven by density wave instability.

Single crystal samples of Na$_2$Ti$_2$As$_2$O used in this study were grown by the NaAs flux method \cite{Shi}. Figure 1 shows the in-plane temperature dependent resistivity and magnetic susceptibility measured in a Quantum Design physical property measurement system (PPMS) and a superconducting quantum interference device vibrating sample magnetometer (SQUID-VSM), respectively. Clear anomalies are observed near 320 K where the resistivity displays a sudden increase and the magnetic susceptibility a sharp drop. Below the phase transition, the resistivity keeps increasing with decreasing temperature, reaches a maximum near 150 K, then decreases slightly. But below roughly 50 K, the resistivity goes up sharply and the compound becomes semiconducting like in the lowest temperature. Such a behavior has not been observed in any other member of the titanium oxypnictide family. In the magnetic susceptibility curve, a weak kink is seen near 42 K, suggesting the emergence of another phase transition. Those anomalies were repeatedly observed in different samples. As will be seen below, optical spectroscopy measurement revealed clearly two distinct DW phase transitions.

\begin{figure}[t]
\centering
\includegraphics[width=2.7 in]{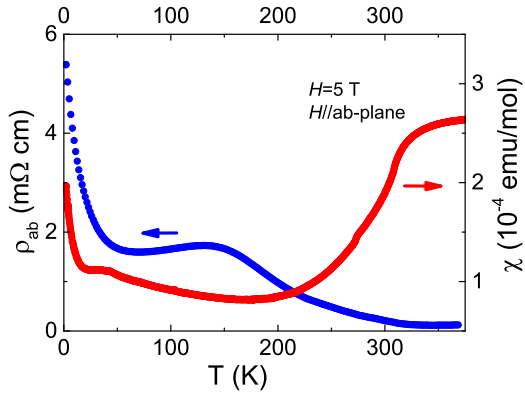}
\caption{(Color online) The temperature dependent in-plane resistivity and magnetic susceptibility of Na$_2$Ti$_2$As$_2$O.}
\end{figure}

Figure 2 and 3 show the reflectance R($\omega$) and real part of conductivity $\sigma_1(\omega)$ spectra at different temperatures. The insets show R($\omega$) and $\sigma_1(\omega)$ spectra in a linear frequency scale. The measurement technique is the same as we presented before \cite{Yue-NTSO}. At 330 K, a temperature higher than the phase transition, R($\omega$) shows a typical metallic response. R($\omega$) approaches unity at zero frequency and drops almost linearly with increasing frequency. This well-known overdamped behavior indicates rather strong carrier scattering. A striking feature is that R($\omega$) is strongly suppressed below 2600 \cm below the phase transition temperature at T$_1$=320 K. Accordingly, a remarkable peak structure emerges in $\sigma_1(\omega)$ and becomes more pronounced upon cooling (left inset of Fig. 3). It is a typical DW energy gap character in the optical conductivity spectrum, which is determined by the "type-I coherent factor" for DW order \cite{Hu-BaFeAs,electrondynamic}. The spectral change across the phase transition is rather similar to that seen for its sister compound Na$_2$Ti$_2$Sb$_2$O, where the phase transition occurs at 114 K \cite{Yue-NTSO}. The peak position in $\sigma_1(\omega)$ roughly gives the gap value of $2\Delta_1$=2600 \cm ($\sim0.32 eV$). This leads to the ratio of the energy gap relative to the transition temperature
2$\Delta_1/k_BT_1\approx$11.5 at 10 K. The value is also close to the that obtained for Na$_2$Ti$_2$Sb$_2$O compound \cite{Yue-NTSO}. Notably, this value is much larger than
the BCS (Bardeen-Cooper-Schrieffer) mean field value of 3.52 for a density wave phase transition \cite{densitywave}. This
means that the transition temperature is significantly lower than the mean field transition
temperature. It might be due to highly two-dimensional electronic structure
which leads to strong fluctuation effect and suppresses the actual ordering temperature. The 2D nature of the compound is well supported by the large value of the anisotropic ratio of the resistivities $\rho_c(T)/\rho_{ab}(T)\sim400$ \cite{Shi}. Similar gap ratios
were also observed in other low-dimensional density wave materials,
e.g. (TaSe$_4$)$_2$I \cite{densitywave}. The right inset of Fig. 3 shows the plot of the energy gap $\Delta_1$, extracted from the optical measurement, as a function of temperature. The data points follow the curve of BCS theory for a density wave phase transition, implying that the transition is a second order phase transition.
It should be emphasized that the low-$\omega$ R($\omega$) in the DW ordered state increases faster towards unity at zero frequency than that at 330 K. As a consequence, one can see a rather sharp low-$\omega$ reflectance edge. This indicates clearly that the Fermi surfaces are only partially gapped and the compounds are still metallic below \emph{T}$_s$.

\begin{figure}
\centering
\includegraphics[width=3.3 in]{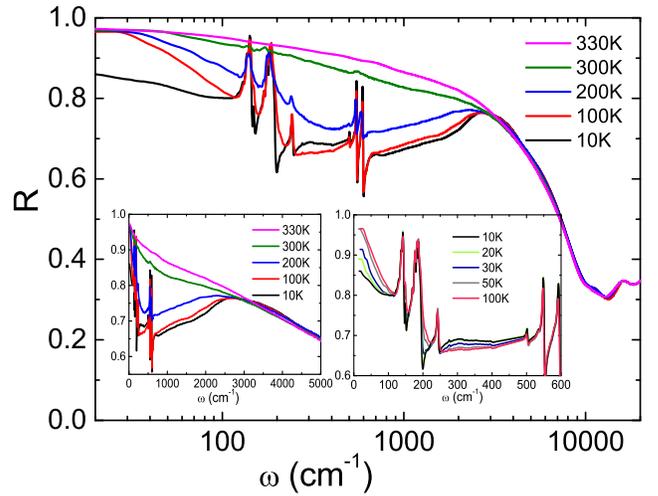}
\caption{(Color online)  R($\omega$) spectra for Na$_2$Ti$_2$As$_2$O at several representative temperatures. left inset:  R($\omega$) spectra up to 5000 \cm in a linear frequency scale; Right inset: R($\omega$) spectra up to 600 \cm at low temperatures.}
\end{figure}

Surprising results were observed at low temperatures. As seen in the right inset of Fig. 2, below 50 K, a new suppression feature appears in R($\omega$) at much lower energy scale, roughly below 300 \cm. Above this energy level, there is a slight enhancement of reflectance. This character becomes more pronounced as the temperature decreases. It is very similar to the spectral change arising from the opening of DW energy gap as we have explained for the phase transition at 320 K. In the $\sigma_1(\omega)$ spectra, the suppression of conductivity at low frequency could be clearly seen below 50 K. Nevertheless, due to the emergence of strong phonon modes, the peak position could not be unambiguously resolved. Overall, this second gap feature is much weaker than the first one at 320 K. This could be attributed to the small residual carrier density being left after the first DW phase transition. Judging from the suppression frequency in R($\omega$) and the spectral weight analysis as being presented below, the gap value is estimated to be about $2\Delta_2$=300 \cm. Combined with the anomalies in resistivity and magnetic susceptibility at $T_2$=42 K, the observation yields compelling evidence for the development of a new DW order. From the gap value, we obtained the ratio of the energy gap relative to the transition temperature $2\Delta_2/k_{B}T_2\sim$10 for the second transition, which is similar to the value for the first DW phase transition. It deserves to remark that the earlier measurement of anisotropic ratio of resisitivities by Montegomery technique on Na$_2$Ti$_2$As$_2$O showed a very sharp drop below 50 K \cite{Shi}, implying a reconstruction of band structure at low temperature, while no such abrupt change in anisotropic resistivities was seen in Na$_2$Ti$_2$Sb$_2$O crystals. Those anisotropic resistivity data also support the presence of the second phase transition below 50 K.

\begin{figure}
\centering
\includegraphics[width=3.3 in]{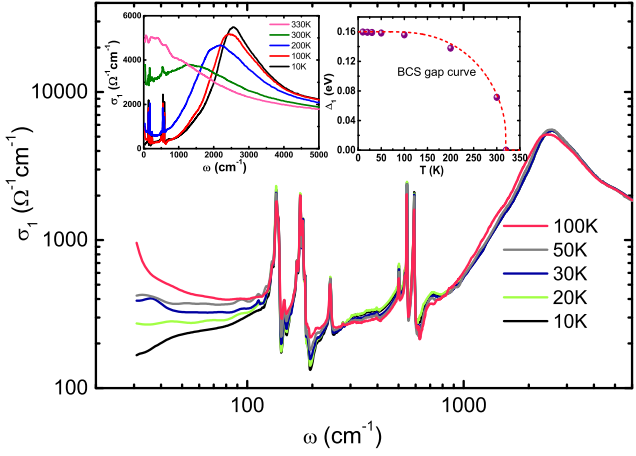}
\caption{(Color online)  $\sigma_1(\omega)$ spectral for Na$_2$Ti$_2$As$_2$O at low temperature over broad range of frequencies. Left inset: The conductivity spectra up to 5000 \cm in a linear frequency scale. Right inset: the plot of energy gap $\Delta_1$, extracted from the optical measurement, as a function of temperature. The data points follow the curve of BCS theory for a density wave phase transition (the dashed curve).}
\end{figure}

At 10 K, the lowest measurement temperature, the reflectance value is finite and smaller than unity at the zero frequency limit. In $\sigma_1(\omega)$, the Drude component is absent and phonon structures become dominant. The very small low-frequency spectral weight is likely due to the effect of finite measurement temperature (10 K). The results indicate that, in the ground state, the Fermi surfaces would be fully gapped and the compound becomes a semiconductor. In 2D DW materials, one usually observes enhanced metallic behavior below the transition because the partial DW energy gap tends to remove the electrons that experience stronger scattering \cite{Hu-TaS2}. Although there are a few examples of nonmetallic 2D CDW systems, the origin of the semiconducting behavior is actually not due to true FS nesting but Mott physics \cite{Colonna,Kim,CaoG}. LaTe$_2$ was suggested to be a 2D FS nesting driven semiconducting CDW system \cite{Carcia}, however, more recent study indicated that the nonmetallic behavior was in fact caused by the localization effect due to the presence of defects in conducting Te layers \cite{Yue-LaTe2}. In this regard, Na$_2$Ti$_2$As$_2$O may be the first example of FS nesting driven semiconducting material.

\begin{table*}[htbp]
\begin{center}
\newsavebox{\tablebox}
\begin{lrbox}{\tablebox}
\begin{tabular}{*{14}{m{8mm}}}
\hline \hline\
{}&$\omega_{p}$&$\gamma_{D}$&$\omega_1$&$\gamma_1$&$S_1$&$\omega_2$&$\gamma_2$&$S_2$&$\omega_3$&$\gamma_3$&$S_3$\\[4pt]
\hline
330K&23&2&--&--&--&--&--&--&5&8&23\\[4pt]
300K&13&1&--&--&--&1.2&2.5&20&5.4&8&23\\[4pt]
200K&4.1&0.3&--&--&--&2.2&1.9&22&5.4&7.5&23\\[4pt]
100K&2.4&0.1&--&--&--&2.5&1.8&22.7&5.4&7.5&23\\[4pt]
50K&2.3&0.09&--&--&--&2.6&1.6&22.5&5.4&7.5&23\\[4pt]
10K&--&--&0.3&0.44&2.6&2.6&1.5&21.5&5.4&7.5&23\\[4pt]

\hline \hline
\end{tabular}
\end{lrbox}

\caption{Temperature dependence of the plasma frequency $\omega_p$
and scattering rate $\gamma_D$=1/$\tau_D$ of the Drude term, the
resonance frequency $\omega_i$, the width $\gamma_i$=1/$\tau_i$
and the square root of the oscillator strength $S_i$ of the
Lorentz component(all entries in 10$^3$ \cm). One Drude mode is
employed at high temperatures. Lorentz terms
responsible for the DW orders are added at low temperatures. The
lowest energy inter-band transition is also displayed.}
\scalebox{1.0}{\usebox{\tablebox}}
\end{center}
\end{table*}

To analyze the evolution of both the itinerant carriers and the density wave gap excitations in a quantitative way, we use a Drude-Lorentz mode to fit the $\sigma_1(\omega)$ spectra in a broad frequency range \cite{Hu-BaFeAs}:
\begin{equation}
\epsilon(\omega)=\epsilon_\infty-{{\omega_p^2}\over{\omega^2+i\omega/\tau_D}}+\sum_{i=1}^N{{S_i^2}\over{\omega_i^2-\omega^2-i\omega/\tau_i}}.
\label{chik}
\end{equation}
Here, the $\epsilon_\infty$ is the dielectric constant at high energy, the middle and last terms are the Drude and Lorentz components, respectively. The Drude term captures the contribution from free carriers and the Lorentz components describe the excitations across the gap and inter-band transitions. It is found that the experimental data below 5000 \cm at 330 K could be reproduced approximately by one Drude term and one Lorentz component.  After the first phase transition, another Lorentz peak (DW gap 1) is added. As the temperature is lowered below 42 K, a third Lorentz term (DW gap 2) develops and takes place the Drude component completely. A list of fitting parameters is displayed in table I. The fitting results for three representative temperatures are shown in Fig. 4. It illustrates clearly that the compound undergoes two phase transitions with different energy gap scales.

It is worth noting that the plasma frequency at 330 K (above the first phase transition) is about 23000 \cm which is almost equal to Na$_2$Ti$_2$Sb$_2$O, while $1/\tau\approx$2000 \cm is much lager than Na$_2$Ti$_2$Sb$_2$O \cite{Yue-NTSO}. This explains the lager resistivity values of Na$_2$Ti$_2$As$_2$O compound. The variations of plasma frequency and scattering rate $1/\tau$ normalized to the values at 330 K as a function of temperature are displayed in the right bottom panel of Fig. 4. Provided that the effective mass of itinerant carriers does not change with temperature, the residual carrier density at 100 K is only 10\% of that at high temperature. Meanwhile, the scattering rate also reduces by 90\%. It indicates that the opening of partial density wave gap remove the electrons near Fermi level which bear stronger scattering. This character is similar to Na$_2$Ti$_2$Sb$_2$O, but the scattering rate of Na$_2$Ti$_2$As$_2$O is less reduced. Those evolutions agree well with the temperature dependence of resistivity. The itinerant carrier density becomes zero at the lowest temperature suggesting the residual Fermi Surface has been fully gapped by the second DW order.

\begin{figure}[t]
\centering
\includegraphics[width=3.3 in]{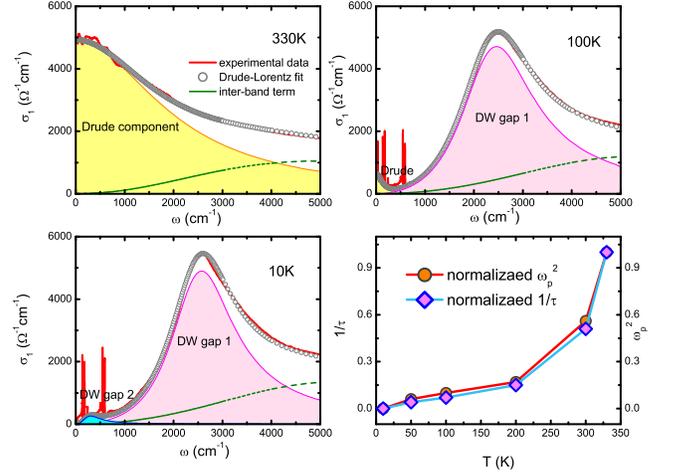}
\caption{(Color online) The experimental data of $\sigma_1(\omega)$ at 330 K, 100 K and 10 K together with the Drude-Lorentz fits; Right bottom panel: The variations of $1/\tau$ and $\omega _{p}^{2}$ normalized to the values at 330 K as a function of temperature.}
\end{figure}

A spectral weight (SW) analysis gives more insight into the development of the two orders. The spectral weight shown in Fig. 5 is defined as $W_{s}=\int_{0}^{\omega _{c}}\sigma _{1}(\omega )d\omega$, where $\omega_c$ is a cutoff frequency. The small step structures below 200 K in low frequency region, between 100 and 600 \cm, are ascribed to the phonons. At very low energy, the $W_{s}$ at 330 K has highest value due to the large Drude component. Below the 300 K , the spectral weight is severely suppressed at about 2000 \cm and recovered at a higher energy. This indicates the formation of an energy gap. Moreover, another suppression at low-$\omega$ region becomes eminent at 20 and 10 K. The SW approaches approximately to zero at finite frequency at 10 K, indicating the insulator character. The suppression is an indication of the second energy gap opening which leads to the M-I transition. We note that the spectral curvatures of $W_{s}$ near 200 - 300 \cm at 10 and 20 K are rather similar to the those near 2000 - 3000 \cm at 100 and 200 K, as indicated by two arrows in Fig. 5. Therefore, the temperature dependence of the SW suppression reflects two different DW energy gap openings below 320 and 42 K, respectively.

\begin{figure}[t]
\centering
\includegraphics[width=2.7 in]{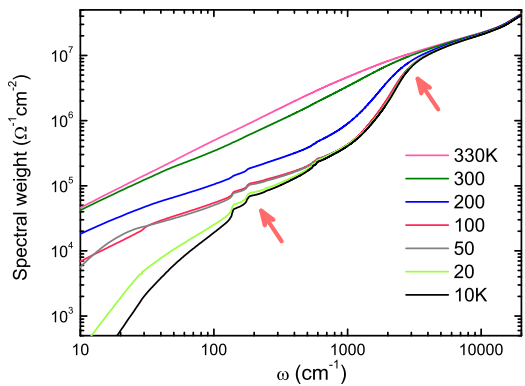}
\caption{(Color online) Frequency-dependent spectral weight at different temperature. The curvatures indicated by arrows reflect the effect of spectral weight redistributions caused by DW gap openings.}
\end{figure}

The observation of two DW phase transitions is totally unexpected. Although a FS nesting driven SDW semiconducting ground state is obtained from the first principle calculations \cite{Yan}, the experimental observations differ in two aspects. First, the two steps phase transitions were not anticipated from the calculations. Second, the value of energy gap in the ordered state from first principle calculations (0.15 eV) is only half of the gap value observed for the first DW phase transition, but four times larger than that for the second DW phase transition. Apparently, further theoretical work is needed to resolve the discrepancy. A more realistic approach to the ground state is to consider first a metastable intermediate DW state driven probably by the strongest nesting instability of FSs, and then to examine further possible instability on the basis of the intermediate DW state.

It also deserves to remark that, although the optical measurement revealed formation of two energy gaps at different temperatures, the measurement could not determine where the FSs are gapped. Momentum resolved experimental probe, such as angle resolved photoemission, should be used to determine the gapped regions and corresponding wave vectors. Another important issue is that, although the optical experiment indicated typical DW phase transitions, the measurement could not tell whether the phase transitions were CDW or SDW since both orders have the same coherent factor and, therefore, the same energy gap character. Other techniques that are capable to probe magnetism should be used to resolve the issue. A recent NMR measurement on BaTi$_2$Sb$_2$O polycrystalline sample revealed an absence of internal field at the Sb site, which therefore favored an CDW origin \cite{NMR}. Nevertheless, this result is actually in agreement with the first principle calculations which indeed predicted a CDW ground state for BaTi$_2$Sb$_2$O \cite{Subedi}. On the contrary, the first principle calculations on Na$_2$Ti$_2$As$_2$O indicated a semiconducting SDW ground state with a blocked AFM order \cite{Yan}. Therefore, sensitive magnetic probes should be applied particularly to Na$_2$Ti$_2$As$_2$O compound. Resolving those issues is of great significance in view of the profound relationship between the density wave (SDW or CDW) and superconductivity in this family. The present study is expected to motivate more experimental and theoretical studies on the materials.

To conclude, we have studied the in-plane optical properties of Na$_2$Ti$_2$As$_2$O single crystal, a sister compound of superconductor BaTi$_2$Sb$_2$O. Unexpectedly, the study revealed two density wave (DW) phase transitions. The first transition at 320 K results in the formation of a large energy gap and removal of most part of the Fermi surfaces. The second phase transition with smaller energy gap scale occurs below 42 K and drives the compound into semiconducting ground state. This makes Na$_2$Ti$_2$As$_2$O the first proven quasi-2D DW semiconductor driven by FS nesting as in 1D materials. The study would shed light on the complex electronic structures of the titanium oxypnictides.

\begin{center}
\small{\textbf{ACKNOWLEDGMENTS}}
\end{center}
The authors acknowledge useful discussions with Z. Y. Lu and J. L. Luo. This work was supported by the National Science Foundation of
China (11120101003, 11327806), and the 973 project of the Ministry of Science and Technology of China (2011CB921701, 2012CB821403).


\begin{references}

\bibitem{densitywave} G. Gr\"{u}ner, \emph{Density Waves in Solids} (Addison-Wesley, Reading, MA, 1994).

\bibitem{Wilson1} J. A. Wilson \emph{et al.},  Adv. Phys. \textbf{18}, 193 (1969).

\bibitem{Wilson2} J. A. Wilson, Phys. Rev. B \textbf{15}, 5748 (1977).

\bibitem{Mongeau} P. Mongeau \emph{ et al.} Phys. Rev. Lett. \textbf{6}, 602 (1976).

\bibitem{Becker} B. Becker \emph{et al.} Phys. Rev. B \textbf{59}, 7266 (1999).

\bibitem{Degiorgi} L. Degiorgi \emph{et al.} Phys. Rev. Lett. \textbf{76}, 3838 (1996).


\bibitem{Yajima}  T. Yajima \emph{et al.}, J. Phys. Soc. Jpn. \textbf{81}, 103706 (2012).

\bibitem{Doan} P. Doan \emph{ et al.}, J. Am. Chem. Soc. \textbf{134}, 16520 (2012).

\bibitem{Zhai} H. F. Zhai Phys. Rev. B \textbf{87}, 100502(R) (2013).

\bibitem{Adam} A. Adam \emph{et al.}, Z. Anorg. Allg. Chem. \textbf{584}, 150 (1990).


\bibitem{Axtell} Axtell III \emph{et al.}, J. Solid State Chem. \textbf{134}, 423 (1997).

\bibitem{Ozawa1} T. C. Ozawa \emph{et al.}, J. Solid State Chem. \textbf{153}, 275 (2000).

\bibitem{Ozawa2} T. C. Ozawa \emph{et al.}, J. E. Chem. Mater. \textbf{13}, 1804 (2001).


\bibitem{Ozawa3}  T. C. Ozawa \emph{et al.}, J. Cryst. Growth \textbf{265}, 571 (2004).

\bibitem{Ozawa4} T. C. Ozawa \emph{et al.}, Sci. Technol. Adv. Mater. \textbf{9}, 033003 (2008).

\bibitem{Liu1} R. H. Liu \emph{et al.}, Phys. Rev. B \textbf{80}, 144516 (2009).

\bibitem{Wang} X. F. Wang \emph{et al.}, J. Phys.: Condens. Matter \textbf{22}, 075702

\bibitem{Liu2} R. H. Liu \emph{et al.}, Chem. Mater. \textbf{22}, 1503 (2010).

\bibitem{Pickett} W. E. Pickett, Phys. Rev. B \textbf{58}, 4335 (1998).

\bibitem{Biani} de Biani \emph{et al.}, Inorg. Chem. \textbf{37}, 5807 (1998).

\bibitem{Singh} D. J. Singh, New J. Phys. \textbf{14}, 123003 (2012).

\bibitem{Yan} X. W. Yan and Z. Y. Lu, J. Phys.: Condens. Matter 25, 365501 (2013).

\bibitem{Subedi} A. Subedi, Phys. Rev. B \textbf{87}, 054506 (2013).

\bibitem{Shi} Y. G. Shi \emph{et al.}, Phys. Rev. B \textbf{88}, 144513 (2013).

\bibitem{Yue-NTSO} Y. Huang \emph{et al.}, Phys. Rev. B \textbf{87}, 100507(R) (2013).

\bibitem{Hu-BaFeAs}  W. Z. Hu Phys. Rev. Lett. \textbf{101}, 257005 (2008).

\bibitem{electrondynamic} M. Dressel \emph{et al.}, \emph{Electrodynamics of Solids: Optical Properties of Electrons in Matter} (Cambridge University Press, Cambridge, UK, 2002).

\bibitem{Hu-TaS2} W. Z. Hu \emph{et al.}, Phys. Rev. B \textbf{76}, 045103 (2007).

\bibitem{Colonna} Colonna \emph{et al.}, Phys. Rev. Lett. \textbf{94}, 036405 (2005).

\bibitem{Kim} Kim \emph{et al.}, Phys. Rev. Lett. \textbf{73}, 2103 (1994).

\bibitem{CaoG} G. Cao \emph{et al.},Phys. Rev. Lett. \textbf{78}, 1751 (1997).


\bibitem{Carcia} D. R. Garcia \emph{et al.}, Phys. Rev. Lett. \textbf{98}, 166403 (2007).

\bibitem{Yue-LaTe2} Y. Huang \emph{et al.}, Phys. Rev. B \textbf{86}, 205123 (2012).


\bibitem{NMR} Kitagawa \emph{et al.}, Phys. Rev. B \textbf{87}, 060510(R) (2013).

\end{references}
\end{document}